\documentclass{elsart}
\usepackage{amsmath,amssymb}
\usepackage{latexsym}
\usepackage{amssymb}
\usepackage{graphicx}
\usepackage{color}
\usepackage{ulem}
\newcommand{\W}{5.5in}
\begin{document}

\begin{frontmatter}
\title{Velocity contrasts enhancement for shear thinning solutions flowing in a rough fracture}
\author[AG,FR]{H. Auradou}
\author[AG,FR]{A. Boschan}
\author[AG]{R. Chertcoff}
\author[AG]{S. Gabbanelli}
\author[FR]{J.P. Hulin}
\author[AG]{I. Ippolito}
\address[AG]{Grupo de Medios Porosos, Facultad de Ingenier\'{\i}a, Universidad de Buenos
Aires, Paseo Col\'on 850, Argentine.}
\address[FR]{Laboratoire Fluides, Automatique et Syst{\`e}mes
Thermiques, UMR 7608, Universit{\'e}s Pierre et Marie Curie-Paris 6
et Paris-Sud, B{\^a}timent 502, Campus Paris Sud, 91405 Orsay Cedex,
France}
\date{Received: date / Revised version: date}
%
\begin{abstract}
 Flow and transport are studied in transparent model fractures with rough complementary self-affine  walls   with  a relative shear displacement $\vec{u}$. The aperture field is shown to display  long range correlations  perpendicular to $\vec{u}$: for flow in that direction, the width and geometry of the front of a dyed shear-thinning polymer solution displacing a transparent one have been studied as a   function of the fluid rheology and flow rate. The front width increases linearly with distance indicating a convection of the fluids with a low transverse mixing between the flow paths.
 The width also increases with the flow-rate as the fluid rheology shifts from Newtonian
 at low shear rates $\dot \gamma$ towards a shear thinning behaviour at higher  $\dot \gamma$ values.
The width also increases with the polymer concentration at high flow-rates. These results demonstrate the enhancement of the flow velocity contrasts between  different flow channels for shear thinning fluids. The relative widths at low and high $\dot \gamma$ values for different polymer concentrations are well predicted by an analytical model considering the fracture as a set of parallel ducts of constant hydraulic apertures. The overall geometry of the experimental front  geometry is also predicted by the theoretical model  from the aperture map.
\end{abstract}

\end{frontmatter}

\section{Introduction}
Transport and flow in porous media and fractured rocks are encountered in many
engineering fields~\cite{Nas1996} and complex fluids such as polymer
gels or surfactants are often involved. 
Applications include enhanced
oil recovery (EOR), drilling muds and heavy oil recovery. In EOR, for instance,
polymer flooding   reduce viscosity driven instabilities (a polymer 
solution is injected in the reservoir and followed by a water flood).
When these complex fluids have shear thinning properties, experimental flow 
measurements~(see \cite{Perrin2006}) 
 display specific features such as permeability enhancements (or reductions of the
effective viscosity) compared to the case of Newtonian fluids. \\
These effects may be strongly influenced by fractures which are frequently encountered
in many reservoirs and generally display  a broad range of  characteristic length scales.
While it is customary to visualize the fractures
 as parallel plates separated by a constant distance~\cite{Bodin2003},
 this representation is rarely accurate:  fracture wall surfaces
are indeed rough and do not perfectly match~\cite{Vickers1992}. This
creates voids of various size resulting in spatial heterogeneities of
the flow field~\cite{Brown1987,Tsang1989}.\\
The objective of the present work is to analyze experimentally and analytically
these velocity contrasts  for shear thinning solutions flowing in
 transparent models  of single  fractures  with rough walls and to determine their dependence
on the fluid rheology and on the flow velocity. The experiments have been realized in 
a configuration
in which flow is strongly channelized as is frequently the case in actual fractures~\cite{Nas1996}:
 this
will be shown to allow for analytic predictions of the relation between the flow distribution 
and  the apertures and, also, of their dependence on the rheological characteristics
of the fluids used in the experiments.\\
We have sought particularly in this work to reproduce the
 roughness of natural fractured rocks which is characterized
by a broad distribution of the characteristic length scales~\cite{Poon1992}. 
More precisely, these surfaces can  often be considered as
 {\it self-affine}~\cite{Mandelbrot1985},
 this means that they remain
statistically invariant under the scaling transformation:
\begin{equation}
  h (\lambda x, \lambda y) = \lambda^{\zeta} h (x, y),
\end{equation}
where $h (x, y)$ is the surface height and $\zeta$ is the roughness
or self-affine exponent. For most materials including granite,
$\zeta$ is close to $0.8$ \cite{Bouchaud2003} but it is
close to $0.5$ for materials such as sandstone and sintered glass 
beads~\cite{Boffa1998,Ponson2006}. 
Many experiments suggest that $\zeta$ is independent on the orientation of 
profiles measured on the surface with respect to the  direction
 of crack propagation (a slight anisotropy has however been recently 
 observed experimentally on some materials~\cite{Ponson2006b}).\\ 
The rough surfaces used in the present
work are transparent milled plexiglas plates with an isotropic self-affine
geometry of characteristic exponent $\zeta = 0.8$: they allow for optical flow 
observations by means of dyed fluids (practically, a transparent solution is displaced
by a dyed one and the geometry of the front is determined by image analysis).
For each fracture, two such complementary surfaces are realized and match perfectly 
when brought in contact: in the model, both a spacing normal to the mean fracture surface
and a relative lateral shift $\vec u$ are introduced in order to create a mismatch and to
obtain a variable aperture field~\cite{Brown1986}.\\ 
While the surfaces are isotropic, previous 
laboratory measurements and numerical 
investigations~\cite{Boschan2006,Gentier1997,Yeo1998,Drazer2004,Auradou2005} 
show that the lateral
shift introduces an anisotropy of the permeability which is highest in the 
direction perpendicular to $\vec u$.
 More precisely, flow channels perpendicular
 to $\vec u$ and with a length similar to the model appear as
 shown in a previous work~\cite{Auradou2006}. 
 As a result, for flow perpendicular to $\vec{u}$, the overall 
geometry
 of  the displacement front
 of a fluid by another of same rheological properties is well reproduced by modelling the
 fracture as a set of parallel ducts with an hydraulic aperture
 constant along the flow~\cite{Auradou2006}: the present work deals exclusively 
 with this configuration.

The
 fluids used here display at low shear rates $\dot{\gamma}$
a``plateau" domain in which they behave as Newtonian fluids of constant viscosity $\mu$ 
while, at higher shear rates, $\mu$ decreases with $\dot{\gamma}$ following a power law.
Comparing the velocity contrasts between the different flow paths in the two regimes allows one therefore to estimate the influence
of the rheology since the  velocity contrasts should be enhanced in the shear thinning case.
Finally, an analytical model predicting the influence of this Newtonian ``plateau" on the
dependence of the velocity contrasts on $\dot{\gamma}$ will be derived and compared to experimental observations.\\
\section{Experimental procedure}
\subsection{Characteristics of the model fracture}
\label{sec:fracture} 
The model fracture  is made of two
complementary rough self-affine surfaces without contact points: both
surfaces are  obtained from a transparent material by means of a 
milling machine and their size  is $85 \ mm \times 170$ \ mm.  
A detailed description of the procedure is given in~\cite{Boschan2006}:
 a self-affine surface $h(x,y)$  is first generated numerically using the
mid-point algorithm~\cite{Voss1985} with a  self-affine exponent $\zeta = 0.8$ 
as observed in many materials~\cite{Bouchaud2003}. A
second surface, complementary from the first one, is generated and
then shifted numerically parallel to its mean plane by $0.33$\ mm.
The milling tool is computer controlled and a complex tortuous path
may be imposed  to obtain the self-affine geometry. 
Moreover, the borders of two parallel sides of the surfaces
rise above the rough surface: they are designed so that, when they
are clamped against the matching border of the other surface, there
is a void space in the remaining areas. The mean planes of
the facing surfaces are parallel outside these borders with a mean distance:
  $\bar{a} = 0.77$\ mm.\\
The local aperture $a(x,y)$ at a location $(x, y)$ in the fracture plane may 
be predicted from the mathematical surface $h(x,y)$ by the relation:
\begin{equation}
  a (x,y) = h (x,y) - h (x,y+u) + \bar{a},
\label{eq:a}
\end{equation}
where $u$ is the lateral shift. 
\begin{figure}[htbp]
\includegraphics[width=\W]{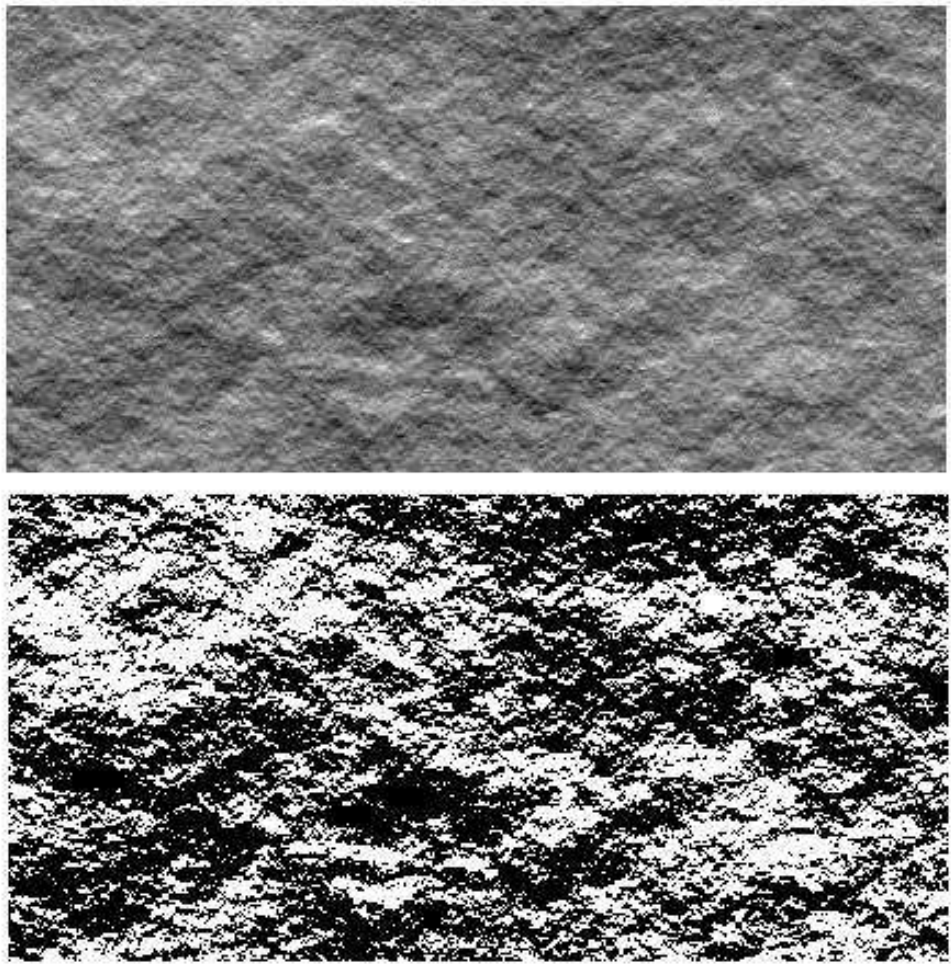}
  \caption
{Upper figure, gray scale representation of the aperture field. Field of view: 
  $85\ mm \times 171$ \ mm. Aperture field - mean value:  $\bar{a} = \langle a (x, y) \rangle_{(x, y)}\, =\, 0.77$\ mm;
  and the aperture fluctuation: $\sigma_a = \langle (a (x, y) - \bar{a})^2 \rangle ^{1 /2}_{(x, y)}\,=\,0.1$\  mm. 
Shift amplitude:  $u\, =\, 0.33$ \ mm (oriented vertically on figure). In the present work, flow is parallel to $x$ direction (horizontal on the figure). 
  Lower image: binarized aperture field with a threshold value equal to the mean aperture ($0.77$ \, mm).}
  \label{fig1}
\end{figure}
Figure~\ref{fig1} shows the
aperture field of the fracture considered in this work:  the binarized
image (lower part of Fig.\ref{fig1}) displays
a clear anisotropy and a large correlation length perpendicular to the shift $\vec{u}$. 
Quantitatively, this effect may be characterized by the
following correlation function, also called semivariance~\cite{Kitanidis1997}:
\begin{equation}
  \gamma ( \vec{\delta}) = \langle (a ( \vec{r}) - a ( \vec{r} +
  \vec{\delta}))^2 \rangle, \label{eq:delta}
\end{equation}
measuring the spatial correlation of the aperture field between
two points separated by a lag vector $\vec{\delta}$. Orientations 
of $\vec{\delta}$ perpendicular  ($x-direction$) and parallel ($y-direction$) 
to the shift are of special interest. 
\begin{figure}[htbp]
\includegraphics[width=\W]{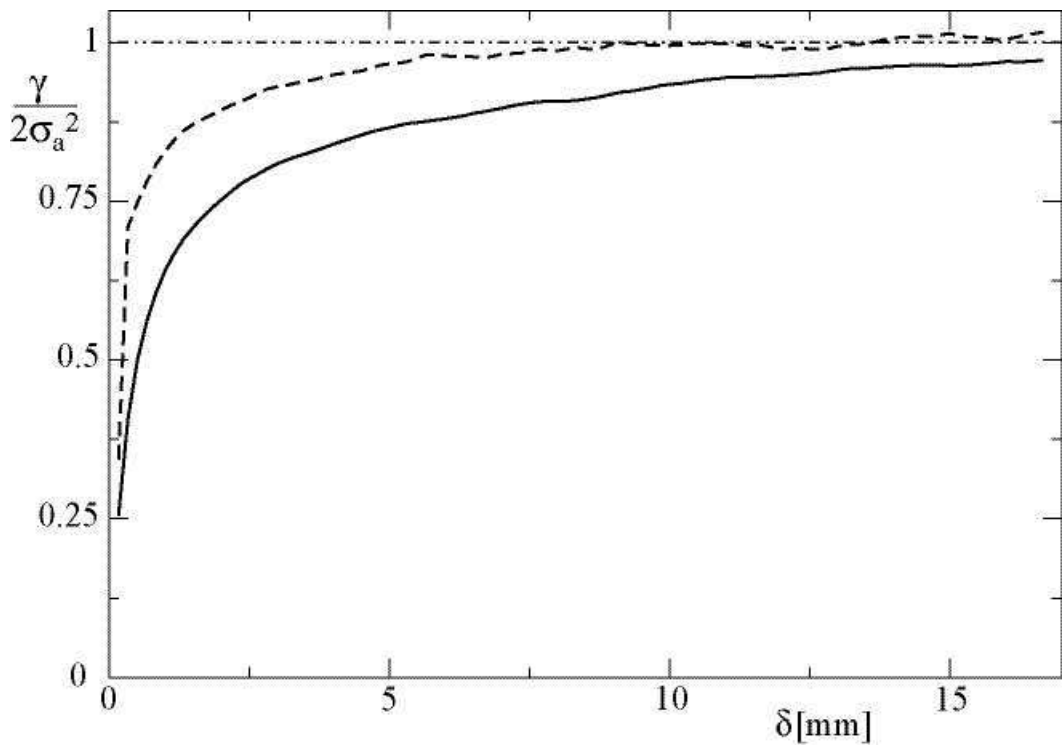}
  \caption{Semivariograms normalized by $2 \sigma_a^2$ as a function of
  lag distance $||\vec{\delta}||$ (mm) for the aperture field displayed in Fig.\ref{fig1}. 
  Dotted line: correlation along the direction $y$ of the shear. Solid line: correlation along 
 the  perpendicular direction $x$ (parallel to the flow  in the rest of the paper).} \label{fig2}
\end{figure}
Figure~\ref{fig2} displays 
variations of the semivariance in both  directions. 
When the lag modulus $||\vec{\delta}||$ is larger than the correlation
 length of the aperture field, one expects
$\gamma$ to reach a constant value equal to $2 \sigma_a^2$, where
$\sigma_a^2=\langle a(x,y)-\bar{a} \rangle ^2$ is the variance of the aperture. 
The semivariance
 $\gamma$ reaches this limit, but in a very different way  for
the two orientations of $\vec{\delta}$. In the direction $y$ parallel to the shift, $\gamma$
becomes of the order of (and sometimes larger than) $2 \sigma_a^2$
 for $||\vec{\delta}|| > 8$\ mm. In the perpendicular direction $x$, $\gamma$ 
 never exceeds the saturation value and slowly increases towards it:
these differences reflect the large scale anisotropic structure of the aperture field. 
Semivariograms have been computed on surface maps of epoxy casts of 
a fractured granite sample in a previous work~\cite{Auradou2006} and display similar features: 
moreover, normalized curves  $\gamma/(2 \sigma_a^2)$ 
corresponding to  different values of $u$ 
displayed a universal variation as a function of the normalized lag $\delta /u$. This suggests
that results obtained in the present work might be extrapolated to other values of $u$.\\
Finally it should be noted that the ratio $S$ of the standard
deviation of the aperture $\sigma_a$ 
to the mean aperture ${\bar a}$
 is only of $0.13$ (Fig.~\ref{fig1}). This implies, as discussed by \cite{Zimmerman1991},
that the fracture can be considered as "hydraulically" smooth with 
 relatively small velocity contrasts between and along flow lines. 
 This keeps  the trajectories of the preferential flow channels 
 relatively straight and  simplifies subsequent analysis.\\
\subsection{Experimental set-up and procedures}
The plexiglas model fracture is held vertically in a fixed position
between a light panel and a $12$ bits digital CCD camera with 
a high stability and dynamical range. Flow
is induced by sucking a dyed solution from the top side while the
lower side is slightly dipped into a bath containing a clear
fluid. An appropriate calibration, described in reference~\cite{Boschan2006},
allows one to obtain from all pictures of each experiment 
 the corresponding concentration map $c(x,y,t)$.
Here, we focus on the geometry of the iso concentration front $c/
c_0 = 0.5$ which is determined by thresholding the
concentration maps and which depends strongly on  the flow heterogeneity.\\
\subsection{Rheological characteristics of shear-thinning soltions} \label{rheosolution}
\begin{figure}[htbp]
\includegraphics[width=\W]{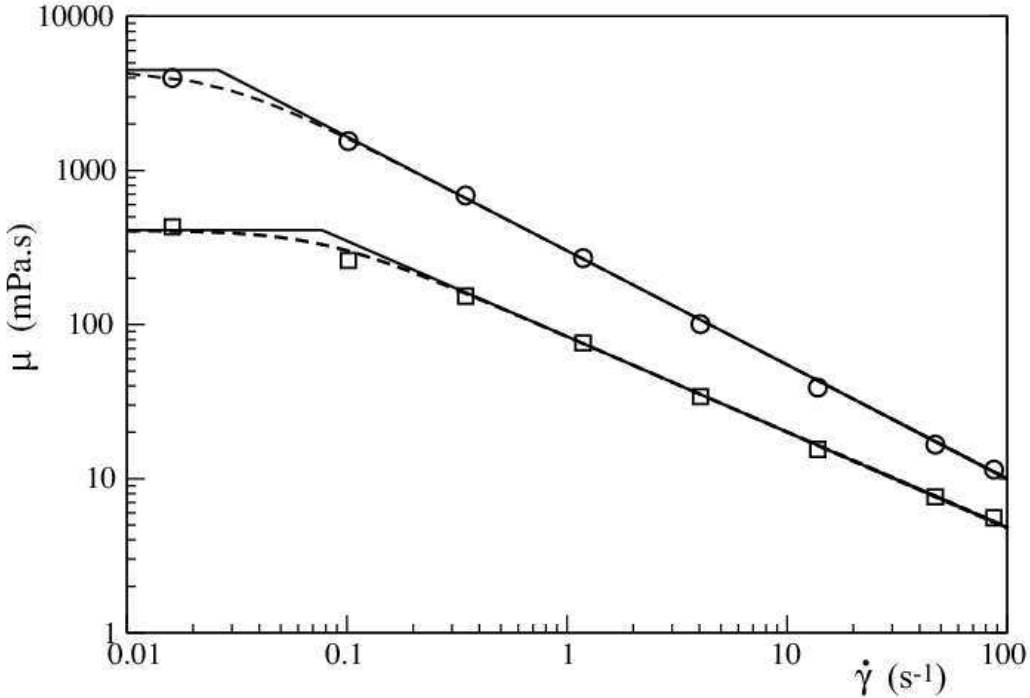}
\caption{Variation of the effective viscosity $\mu$ of the polymer solutions as a 
function of the shear rate $\dot \gamma$ for two water-polymer solutions of different concentrations: $500$\ ppm ($\square$) and  $1000$\  ppm ($\circ$).  Dashed lines: Carreau functions
corresponding to the sets of parameters of Table~\ref{tab1}; continuous lines:
truncated power law approximation.}
\label{fig:fig3}
\end{figure}
In this work, we used shear thinning polymer solutions, more specifically 
water-scleroglucan solutions; they have been characterized using
a {\it{Contraves LS30}} Couette rheometer for shear rates
$\dot{\gamma}$ ranging from $0.016 \hspace{0.25em}$ \ s$^{- 1}$ up to
$87 \hspace{0.25em}$\ s$^{- 1}$. Two different polymer
concentrations equal to
 $500$ \ ppm and $1000$ \ ppm have been used. 
The rheological
properties of the dyed and transparent solutions have been verified to be constant with
time within experimental error over $3$ days;  the variation of the effective viscosity $\mu$
as a function of the shear rate $\dot{\gamma}$ is displayed in Figure~\ref{fig:fig3}. The
variation of $\mu$ with $\dot{\gamma}$ is well
adjusted by a Carreau function:
\begin{equation}
    \mu = \frac{1}{(1 + (
  \frac{\dot{\gamma}}{\dot{\gamma_0}})^2)^{\frac{1 - n}{2}}} (\mu_0 -
  \mu_{\infty}) + \mu_{\infty} .
  \label{powervisc}
\end{equation}
The values of the corresponding rheological parameters for the polymer solutions
characterized in the present work are listed in Table~\ref{tab1}.
$\mu_{\infty}$ is too low to be determined within the available range of 
shear rates ($\dot \gamma \le 87$\ s$^{- 1}$) and it has been taken equal
to the viscosity of the solvent ({\it i.e.} water). In Eq.(\ref{powervisc}), 
$\dot{\gamma_0}$ corresponds to a crossover
between two behaviors. On the one hand, for $\dot{\gamma} < \dot{\gamma_0}$, the
viscosity $\mu$ tends towards the limiting value $\mu_0$, 
and the fluid behaves as a
Newtonian fluid. On the other hand, if $\dot{\gamma} > \dot{\gamma_0}$,  
the viscosity follows a power law variation reflecting its shear thinning characteristics
with  $\mu \propto \dot{\gamma}^{(n - 1)}$.
\begin{table}[htbp]
\begin{center}
  \begin{tabular}{lccc}
    Polymer Conc.  &  $n$  &   $\dot{\gamma_0}$  &   $\mu_0$\\
    ppm  &   &  $s^{- 1}$  &   $mPa.s$\\
    &   &   &  \\
        $1000$  &  $0.26 \pm 0.02$ &   $0.026 \pm 0.004$  &   $4490 \pm
        342$\\
    $500$  &  $0.38 \pm 0.04$  &  $0.077 \pm 0.018$  &  $410 \pm 33$
\end{tabular}
  \caption{Rheological parameters of schleroglucan solutions used in the flow  experiments.}
  \label{tab1}
\end{center}
\end{table}
For each experiment, the flow rate is kept constant at a value  between $0.01$\ ml/min 
and $5$\ ml/min (corresponding mean flow velocities: 
$0.0003 \le v \le 0.14$  mm.s$^{-1}$). Under such conditions the
typical shear rate $\dot{\gamma} \simeq v / a$
ranges between $4.10^{- 4}$\ s$^{- 1}$ and $0.18$\ s$^{- 1}$.
 The latter value is far below the shear rate corresponding to
the second Newtonian plateau ($\mu = \mu_{\infty}$) and this limit will not be
considered in this work. On the contrary, the lowest
 values of the typical shear rate are much lower than $\dot{\gamma_0}$:
  the Newtonian ``plateau" in the rheological curves may therefore have
  a crucial influence of the flow properties.

In order to obtain an analytical expression accounting for the effect of the
fluid rheology on the velocity fluctuations,  the rheological law of the fluids is
approximated in section~\ref{modelization} by a truncated power law.
  
When $\dot{\gamma} < \dot{\gamma_0}$,  the viscosity 
$\mu ( \dot{\gamma})$
is considered 
as constant and equal to $\mu_0$; for   $\dot{\gamma} > \dot{\gamma}_0$,  
$\mu ( \dot{\gamma})$
 is assumed to follow a  power law  $\mu (\dot{\gamma}) = m \dot{\gamma}^{n - 1}$
in which  $m=\mu_0 / \dot{\gamma_0}^{n-1}$. The parameters $n$, $\dot{\gamma}_0$
and $\mu_0$ are obtained from Tab.\ref{tab1}.\\

\section{Flow velocity dependence of front geometry}
\begin{figure}[htbp]
\includegraphics[width=\W]{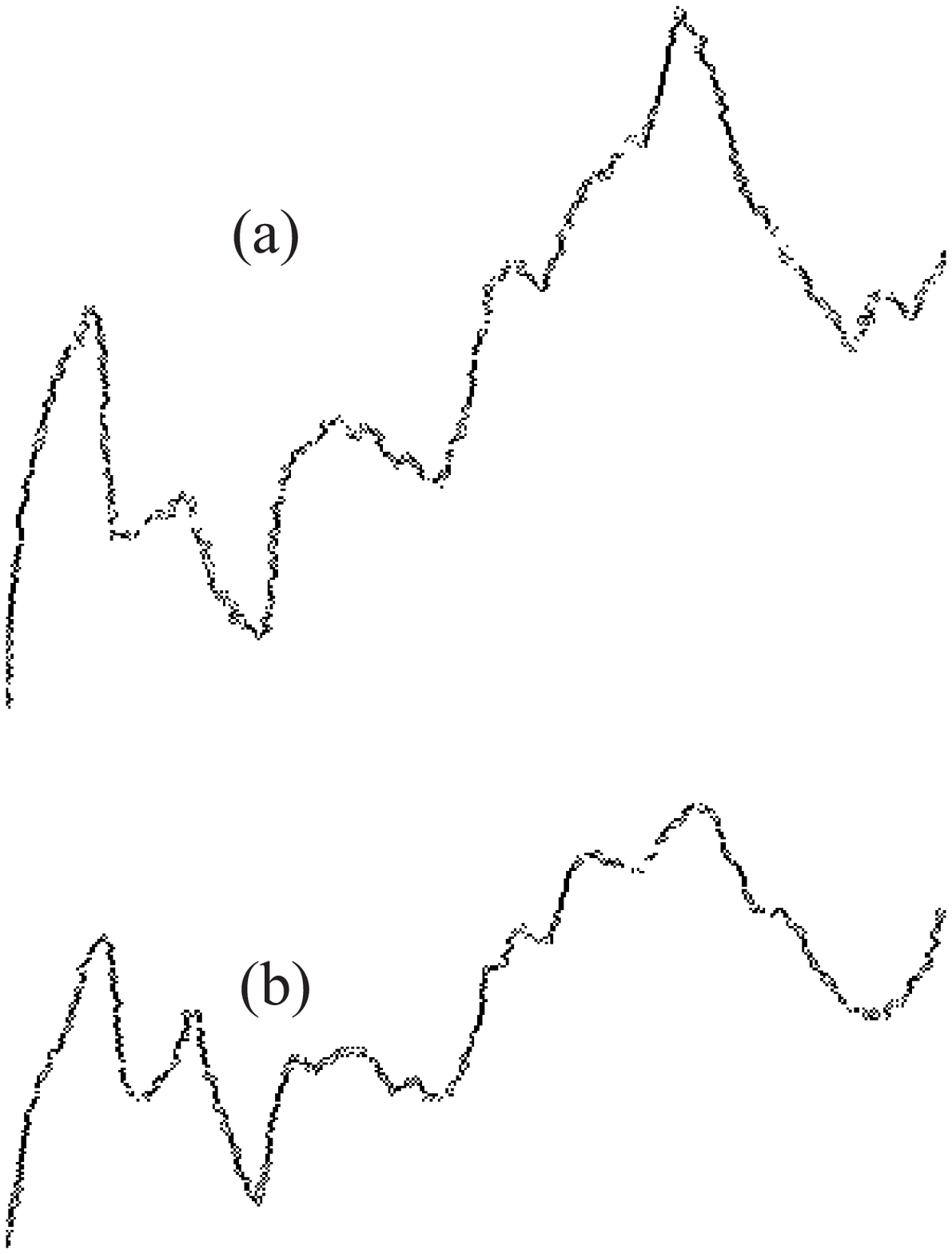}
  \caption{Displacement fronts  at 
  two different flow rates. (a) $Q=1$\ ml/min 
($\bar{v}/v_c \simeq 100$); (b) $Q=0.1$\ ml/min ($\bar{v}/v_c \simeq 10$).
 Polymer concentration: $1000$\ ppm. Vertical mean flow. 
 Front width  perpendicular to mean flow: $85$\ mm. 
 Front extension along the flow: $53$\ mm (top curve), $34$\ mm (bottom curve).
 Mean distance of the front from the inlet equal to half the fracture length.}
\label{fig:fig4}
\end{figure}
\begin{figure}[htbp]
\includegraphics[width=\W]{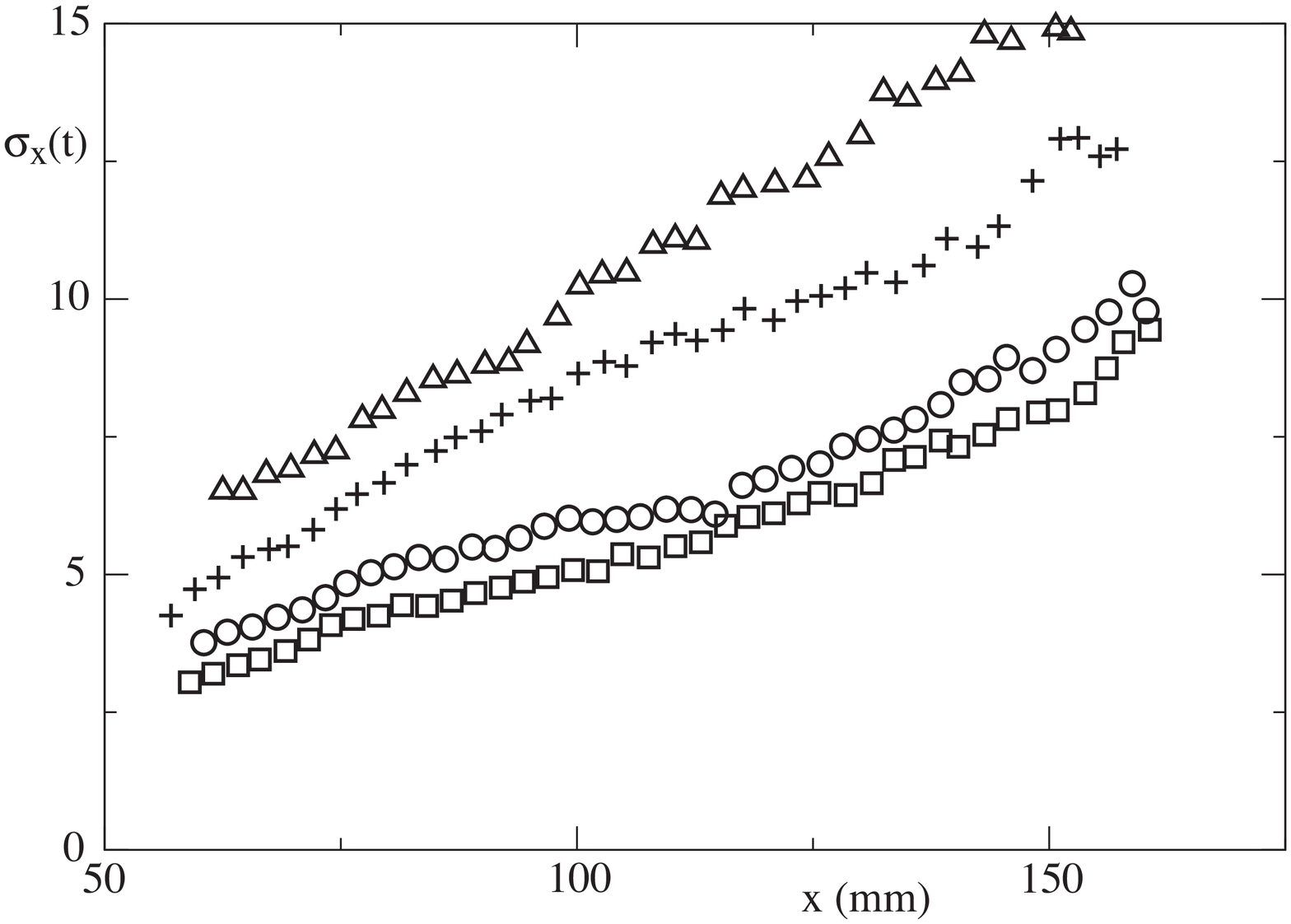}
  \caption{Variation of the mean front width
$\sigma_x(t)=\langle (x(t)-\bar{x}(t))^2 \rangle ^{1/2}$
as a function of the mean distance
$\bar{x}(t)$ from the inlet side of the model. Polymer concentration:  $1000\ ppm$. 
Flow rates:  $Q \, =$ ($\triangle$)$2.0$,($+$)  $0.5$, ($\circ$) $0.1$, ($\Box$) $0.02$\ ml/min.}
\label{fig:fig5}
\end{figure}

Two important features of the displacement front have been observed: (a)
its geometry depends on the flow rate $Q$, and (b) its width
parallel to the flow increases linearly with the distance from the
injection side. The first point is illustrated by Figure~\ref{fig:fig4} in which
two fronts measured are compared during fluid displacements
at two different flow rates, but for a same polymer concentration ($1000$\ ppm): 
the front width parallel
to the flow direction is clearly larger at the highest flow rate.\\
The broadening of the front may be characterized quantitatively from the variation
of the mean square front width,
 $\sigma_x(t)=\langle
(x(t)-\bar{x}(t))^2 \rangle ^{1/2}$,
 as a function of the mean
distance $\bar{x}(t)$ of the front from the injection side (Figure
\ref{fig:fig5}). 
For all values of $Q$, $\sigma_x(t)$ increases linearly with
$\bar{x}(t)$. 
In the next section, this will be shown   to result directly from
 the underlying channelized structure of the aperture field. The width of the front 
after a transit time $t$ corresponds then directly to the product 
$t \Delta v$  where $\Delta v$  is the 
velocity difference between the  different channels (the transverse exchange 
between channels is too small to allow one to reach a diffusive spreading regime).\\
At all distances, the width $\sigma_x(t)$ increases
 with the flow rate $Q$ but with a particularly sharp variation between
 $Q = 0.1$\ ml/min and  $Q = 0.5$\ ml/min.
It will be seen that, at this transition flow-rate, the shear rate at
the fracture walls becomes of the order of $\dot{\gamma_0}$ 
(the threshold value above which the fluids display shear thinning characteristics).\\

\section{Modelization} \label{modelization}
\subsection{Flow between parallel plates with a constant gap}
We compute the flow of the polymer solutions between parallel plates using the
same approach as in reference~\cite{Gabbanelli2005}. 
The relation between the longitudinal pressure drop
and the velocity profile in the gap  is obtained by using the truncated
power law model discussed in section~\ref{rheosolution}. The flow
 field is unidirectional and parallel to $x$ so that $v_x(z)$ is the only 
non zero velocity component. The strain rate is given by: $\dot{\gamma}(z)={dv_x}/{dz}$.\\
At low pressure gradients, the fluids behave
 like a Newtonian fluid with a constant viscosity $\mu_0$  and the resulting
 velocity profile is parabolic and symmetrical between the walls.
Then, the shear rate is zero half way between the fracture walls
and  reaches a maximum at their surface where
$\dot{\gamma} = {6v/a}$ ($a$ is the distance between the plates).                                                                                                                                                                                                                                                                                                                                                              
This value of $\dot\gamma$ is proportional to the mean flow velocity 
or, equivalently, to the pressure drop.\\
As the flow rate keeps increasing,  $\dot{\gamma}$ becomes larger
 than $\dot{\gamma_0}$ and the non Newtonian characteristics of the
  fluid modify the   velocity profile.
The mean flow velocity $v_c$ corresponding to the transition between the two regimes 
satisfies:  $v_c \ =\ a \dot{\gamma_0} / 6$ and the corresponding pressure gradient 
is:  $\nabla P_c = 2\mu_0 \dot{\gamma_0}/a$.\\
As $v$ increases above $v_c$,  the layer  where the shear rate is higher 
than $\dot{\gamma_0}$ becomes thicker and the  velocity profile  $v_x(z)$ is no 
longer parabolic: the full expression may be derived analytically and  is given in Eq.~(5)  
of reference~\cite{Gabbanelli2005}.\\
The mean velocity, $v$, can then be computed by integrating $v_x(z)$ over the
fracture gap, leading to:
\begin{equation}
 \label{eq1}
v = \frac{a^2}{12(2n+1)\mu_0}\nabla P . \left[ (1-n)(\frac{\nabla P}{\nabla P_c})^{-3} 
+ 3n (\frac{\nabla P}{\nabla P_c})^{\frac{1-n}{n}}\right].
\end{equation}
We consider now the case of shear thinning fluids such that $n>0$ and 
 $(1 - n) / n > - 1$.
Then, the leading term in Eq.(\ref{eq1}) is
 $(\nabla P/\nabla P_c)^{(1-n)/n}$ and, therefore, when 
 $\nabla P >> \nabla P_c$, Eq.~(\ref{eq1}) becomes:
\begin{equation}
  \label{eq2} 
v \simeq  \frac{a^2}{12} \left (\frac{\nabla P}{\mu_{eff}} \right )^\frac{1}{n},
\end{equation}
where $\mu_{eff} = \mu_0 \left ( 2 \dot \gamma_0 / a \right )^{1-n}((2n+1)/3n)^n$. 
This is similar to   the generalized  version of
Darcy's law often applied to the flow of non Newtonian 
and, more specifically, to 
power law fluids
in porous media~\cite{Balhoff2006,Shah1995,Fadili2002}.
\subsection{Flow in rough fractures} \label{flowrough}
In this part, we focus on the variations of the velocity in the plane 
$(x,y)$ of  the fracture
and we assume therefore a two-dimensional flow field $\vec{v}(x,y)$ equal
to the average of the fluid velocity profile over the gap 
with $\vec{v}(x,y) = \langle \vec{v}(x,y,z)\rangle _z$.\\
The development with time of the front (represented by the iso concentration lines $c/c_0 = 0.5$)
 will now be analyzed by assuming that its points move at the local
 flow velocity $\vec{v}(x,y)$ and an analytical model predicting the global front 
 width will be developed.\\
This model is based on the results of a previous work
\cite{Auradou2006} 
 demonstrating that, in such systems, the aperture 
field is structured into channels perpendicular to the lateral shift $\vec{u}$
of the surfaces. For a mean
flow parallel to these channels, the paths of the tracer particles
have a weak tortuosity; also, the velocity variations along these
paths are small compared to the velocity contrasts between the
different channels. Under these assumptions, the velocity of a
particle located at a distance $y$, perpendicular to the mean
velocity, satisfies:
\begin{equation}
\vec{v}(x,y) \approx v(y) \vec{e_x},
\end{equation}
where $\vec{e_x}$  is the unit vector parallel to the mean flow. Note
also that, in the geometry discussed in this section, there are no
contact points between the walls of the fractures: this avoids 
 to take into account the large tortuosity of the flow lines in their
vicinity.\\
If the fluid is Newtonian with a constant
viscosity, then, for each channel, the velocity is related to the
pressure gradient $\nabla P$ by relation~(\ref{eq2}) with $n=1$;
 $a$ is now an equivalent (or hydraulic) aperture associated to each
channel and noted $a(y)$ and the equation represents the classical linear
equivalent of Darcy's law for fractures. 
Previous studies have shown that, for relatively small
aperture fluctuations, this hydraulic aperture is well approximated
by the geometrical aperture \cite{Brown1987,Zimmerman1991}: this
suggests that $a(y)^2$ can be taken equal to the mean of the average of 
the square of the local apertures along
the direction $x$ {\it{i.e.}} $a(y)^2\ = \ \langle a(x,y)^2\rangle _x$.
The validity of this assumption has been tested numerically previously for
 a similar geometry~\cite{Auradou2006} in the case of a Newtonian fluid: 
these simulations used  the lattice Boltzmann method to determine the 
$2D$ front geometry at all times: except for fine scale details,  the profile
 $x(y,t)$ of the distance of the front from the inlet at a given time $t$ 
 follows very closely the variations of $a(y)^2$.\\
For a power law fluid such that $n<1$,
the velocity satisfies  the non linear  generalized relation~(\ref{eq2}).
We seek now to generalize to this case the relation between the
front geometry and the aperture variation established for the Newtonian
fluids:  the aperture field is still assumed to be strongly correlated  in the flow
 direction, allowing one to consider the fracture as a set of parallel ducts.
\\
We consider particles starting at $t=0$ from the inlet of the model at different transverse
 distances $y$ and moving at different velocities $v(y)$. Then the
 distance $x$ of the particles from the inlet at time $t$ after the injection satisfies
 $x(y,t)=v(y) t \ $  so that the mean distance of the front from the inlet side is
 $\bar{x}(t)=\langle v(y)\rangle _y t=\bar{v}t$ and:
\begin{equation}
\label{eq0}
\frac{x(y,t)}{\bar{x}(t)} = \frac{v(y)}{\bar{v}}.
\end{equation}
Moreover,  the mean square deviation
$\sigma_x(t)= \langle (x(y,t)-\bar x)^2 \rangle ^{1/2}$ should
satisfy: $\sigma_x(t)=\sigma_v t$ where $\sigma_{v}$ is the
 mean square deviation of the velocities in individual channels from their 
 mean value $\bar{v}$. Combining the previous relations leads to:
\begin{equation}
\frac{\sigma_x (t)}{\bar{x}(t)}=\frac{\sigma_v}{\bar{v}}.
\label{eq:front}
\end{equation}
This equation shows that there is a direct relation between the
front geometry and the variations of the velocity 
from one channel to another: for power law fluids, the latter are 
related to the variations of the hydraulic aperture by Eq.~(\ref{eq2}). 
In order to estimate these variations, we introduce a modified reduced
aperture deviation $S_h$ defined as the ratio between the standard 
deviation of the hydraulic aperture $a(y)$ to its mean. The parameter
 $S_h$ is equivalent to the reduced aperture deviation $S$ defined in
section~\ref{sec:fracture} but the geometrical aperture is replaced 
by the hydraulic one.
Here, we are  interested in weakly fluctuating systems, {\it{i.e.}} for 
which both $S$  and $S_h$ are small compared to one. In addition,  
the hydraulic aperture $a(y)$ is observed to follow a
Gaussian distribution. 
Moreover, Eq.~(\ref{eq2}) shows that, for a given pressure gradient $\nabla P$, 
$v$ scales as $a^{n+1/n}$ : together with the above assumptions, this leads to
the following relation between the 
reduced velocity fluctuations $\sigma_v/ \bar{v}$ and $S_h$:
\begin{equation}
\frac{\sigma_v}{\bar{v}}=\frac{n+1}{n} S_h.
\label{eq:n}
\end{equation}
Combining Eqs.~(\ref{eq:front}) and (\ref{eq:n}), leads to:
\begin{equation}
\frac{\sigma_x(t)}{\bar{x}(t)}=\frac{n+1}{n} S_h.
\label{eq:final}
\end{equation}
\section{Quantitative comparison between the experiments and the model}
In the present experiments, the polymer solutions are expected to behave like
Newtonian fluids as long as the shear rate $\dot{\gamma}$ is
everywhere lower than the critical value $\dot{\gamma_0}$ (see Table \ref{tab1}). As
 the flow rate increases, the critical shear rate $\dot{\gamma_0}$ is first
 reached at  the wall of the fracture where $\dot \gamma$ is highest.
If the fracture is modeled as two parallel plates separated by the
mean aperture $\bar{a}$, then $\dot \gamma = \dot{\gamma_0}$ 
at the walls when the mean flow velocity is 
$v_c=\bar{a} \dot{\gamma _0}/6$. Above this velocity, the shear thinning
properties of the fluids influence the flow and enhance the velocity
fluctuations.

\begin{figure}[htbp]
\includegraphics[width=\W]{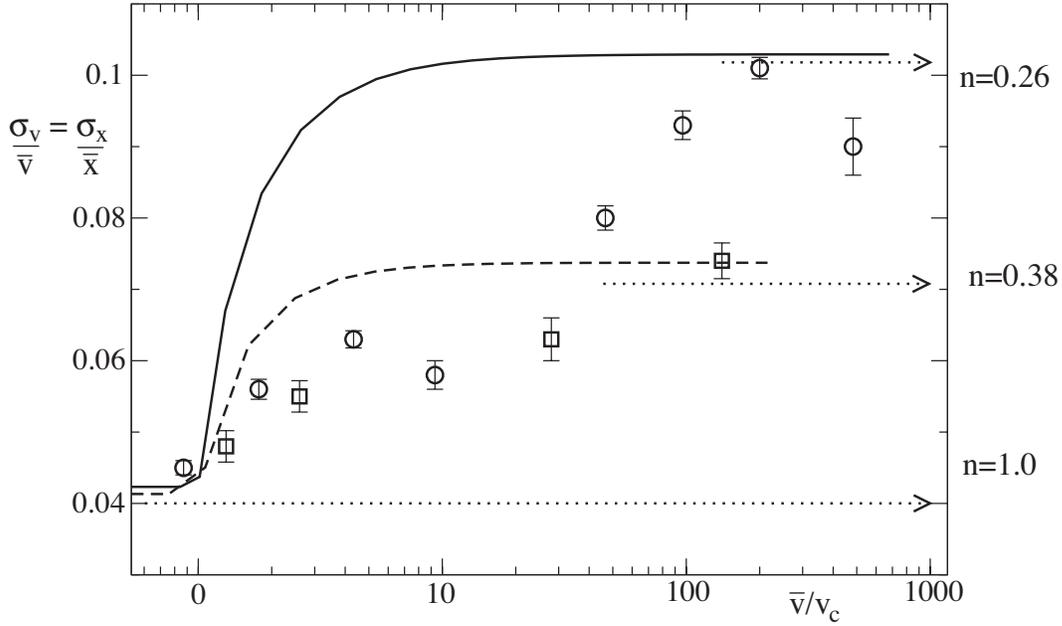}
 \caption{Experimental variation of the ratio
$\sigma _v / \bar{v} = \sigma _x / \bar{x}$  in the model fracture 
($S_h \simeq 0.02$) as a
 function of the normalized mean velocity
$\bar{v} / v_c$ for shear thinning solutions of concentrations $1000
\ ppm$ ($\circ$) and $500 \ ppm$ ($\Box$). Horizontal dotted lines:
theoretical values of $\sigma_v / \bar{v}$ computed
 from Eq.(\ref{eq:final}) for a Newtonian fluid ($n=1$) and for $500 \ ppm$
  (resp. $1000 \ ppm$) polymer solutions ($n=0.38$, resp. $0.26$).
Solid and dashed lines: variations of
$\sigma_v / \bar{v}$ a as function of $\bar{v} / v_c$ computed by
integrating Eq.~(\ref{eq1}.)} 
\label{fig:fig6}
\end{figure}
\begin{figure}[htbp]
\includegraphics[width=\W]{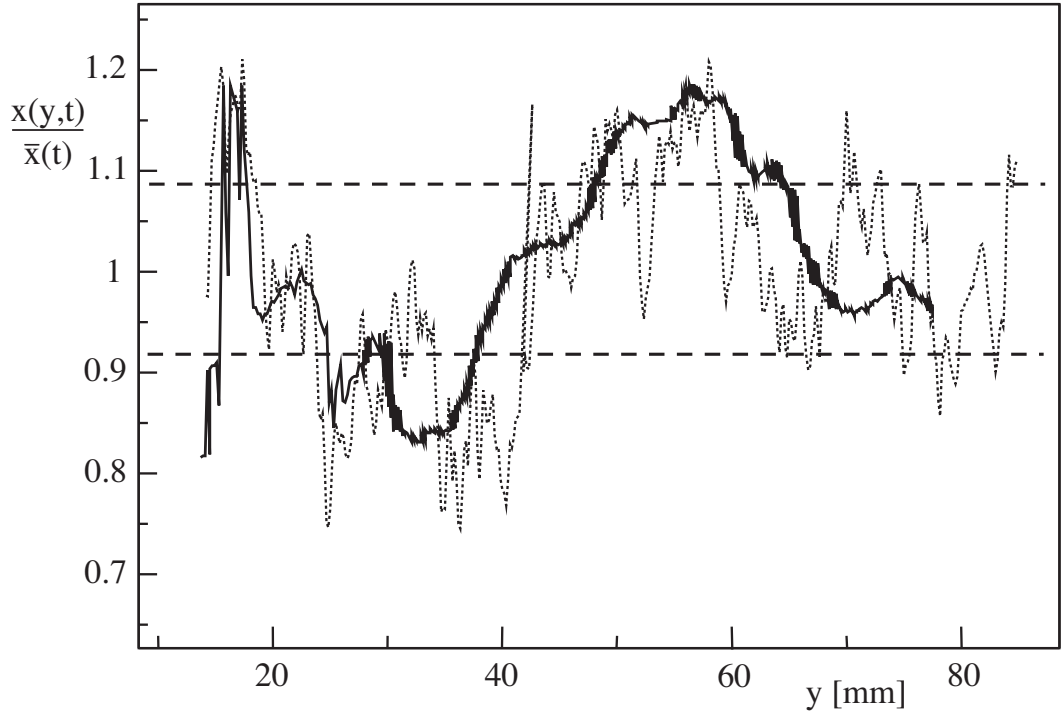}
\caption{Thick solid line: Experimental normalized front profile    $x(y,t)/\bar{x}(t)$ 
as a function of the transverse distance $y$ (mm) for $\bar{v}/v_c=200$ for a
 $1000$\ ppm shear thinning polymer solution. Dotted line: theoretical 
 variation of the normalized velocity $v(y)/\bar{v}$ in the parallel flow channel model.
The front is displayed  just before the displacing fluid starts to
flow out of the fracture. Dashed lines: characteristic deviations of the distance $x(y)$ 
from its mean value $\bar{x}$.} 
\label{fig:fig7}
\end{figure}

Fig.~\ref{fig:fig6}  displays the experimental variations of the normalized
velocity fluctuations (equal to the 
normalized front width $\sigma_x(t)/\bar{x}(t)$) as a function of the
reduced velocity $\bar{v}/v_c$ for both polymer solutions.
The values predicted by Eq.~(\ref{eq:final})
for a Newtonian fluid ($n=1$) and for power law fluids with the
same index as the two solutions are also plotted. \\
For  $\bar{v}/ v_c <1$ the
experimental values are similar for both solutions and close to the
theoretical prediction for $n=1$ (horizontal dashed line). For  $\bar{v}/ v_c \gg 1$,
$\sigma_x(t)/ \bar{x}$ 
tends toward values  of the order of those predicted by Eq.~(\ref{eq:final}) and increasing
with the polymer concentration.\\
 Eq.~(\ref{eq:final}) provides therefore a good
estimate of the velocity fluctuation inside the fracture both for low,
{\it{i.e.}} $\bar{v}/ v_c<1$,  and high flow rates corresponding to
$\bar{v}/ v_c>>1$. The increase of $\sigma_x(t)/ \bar{x}$ between the Newtonian
and shear thinning regimes and also, at high velocities, with the polymer concentration
confirms the enhancement of the velocity contrasts between the channels 
for shear-thinning fluids.\\
Between the limiting values $\bar{v}/v_c < 1$ and $\bar{v}/v_c \gg 1$, fluid velocity
variations within the fracture may be estimated by applying Eq.(\ref{eq1}) in each
channel (assumed to be of constant hydraulic aperture): this equation takes into
 account the coexistence in the fracture gap of layers where the fluid has
 Newtonian and non Newtonian properties. The normalized velocity
fluctuations $\sigma_v/\bar{v}$ obtained by these computation are
displayed in Fig.\ref{fig:fig6} for the two polymer concentrations together with
the experimental variations of the normalized front width $\sigma_x(t)/\bar{x}(t)$.\\
In agreement with the theoretical curves, $\sigma_x(t)/\bar{x}(t)$ starts to increase 
when the velocity $\bar{v}$ becomes larger than $v_c$ 
(${\bar v}/v_c > 1$) for both  polymer solutions.
However, although the limiting value for $\bar{v}/ v_c \gg 1$ is the same as 
predicted,  the increase  of $\sigma_x(t)/\bar{x}(t)$ above $v_c$ is slower
than expected: actually, the theoretical predictions represent an upper bound for
the observations.\\
This difference may be due in part  to the use of a simplified version of the rheological
curve displaying a transition sharper than the actual one between the Newtonian and
shear thinning regimes. Also, the aperture of the  parallel channels introduced in
the model is assumed to be constant: this also leads to a transition between the 
Newtonian and power law regimes which is faster than the actual one.\\
A step further in the interpretation is the comparison of the experimental shape of the
fronts with that estimated
from the channel model. In section~\ref{flowrough},  the normalized distance
$x(y,t)/\bar{x}(t)$ of the front from the inlet has been predicted to be equal to $v(y)/\bar{v}$
(see Eq.~\ref{eq0}). An experimental  front profile  normalized in this way is plotted in
Fig. \ref{fig:fig7} as a function of the transverse distance $y$ together with the variation
of the theoretical normalized velocity $v(y)/\bar{v}$. The velocity $v(y)$ is estimated from
 Eq.~\ref{eq2} in which the aperture $a$ is replaced by the mean value $a(y)$ defined in
  section~\ref{flowrough}.\\
 The most remarkable observation is the fact that both the experimental and 
theoretical fronts have not only the same width but also nearly the same geometry.
These results are very similar to those of numerical simulations
for Newtonian fluids~\cite{Auradou2006}: they demonstrate the validity of the generalization
 in Eq.~(\ref{eq2}) of  the Newtonian model.
Fine scale details  predicted by the theoretical model are however not observed in the
experimental front: this difference may
be due in part to viscous drag forces between parallel
layers of fluid moving at different velocities in the fracture plane.
These forces may smoothen
the local velocity gradients and rub out small scale features of the front 
without changing the large scale velocity variations: this results in a bumpy
front with  a typical width of the structures of the order of $10\ mm$. This latter
value is of the order of the correlation length in the direction perpendicular to the
channels.
\section{Discussion and conclusions} \label{conclusion}
In the  present work the enhancement of 
velocity fluctuations  for shear thinning fluids has been studied in
a single fracture with rough, self-affine walls. The two wall surfaces are perfectly
matched and are positioned with both a normal and a lateral shift.
 This results in an anisotropic aperture field well  characterized quantitatively
 by the semivariograms of the aperture both in the direction of the shift and
 perpendicular to it. The characteristics of these semivariograms are in agreement
with previous experimental measurements on granite samples~\cite{Auradou2006}.
  Parallel to  the shift, the aperture field is correlated over a distance of
the order of $10\ mm$ above which the value of the normalized
semivariograms is of the order of $1$. In the other direction, 
the correlation subsists over the full fracture length. \\
This observation has allowed us to
model the fracture as a set of parallel ducts perpendicular to the shift and with an
hydraulic aperture constant along their length. 
These assumptions lead  to specific predictions on the dependence of the width and
of the geometry of the front on the velocity of the fluid and its rheology: these 
predictions deal with the case of a mean flow parallel to the channels which was the
configuration used in the present experiments.
This model generalizes a previous one developed for  Newtonian fluids and which
has been validated by numerical simulations~\cite{Auradou2006}.\\
The variation of the front width with the velocity could first be predicted. 
At low flow rates, the viscosity of the solutions
is constant (Newtonian ``plateau") but non Newtonian effects become
important for faster flows: this results in an increase of
the velocity fluctuations -and of the front width. This variation occurs when
  the shear rate at the fracture wall becomes larger
$\dot{\gamma_0}$, {\it i.e.} the shear rate corresponding to the crossover
between the Newtonian plateau and the power law regimes: $\dot{\gamma_0}$ is
reached for a mean flow velocity $v_c = a \dot \gamma_0 / 6$. At still higher
flow velocities of the order of  $100 \times v_c$,
both the normalized velocity fluctuations and the normalized front
width reach a new constant value with a good agreement between the experimental
results and the theoretical expectations. \\
These results validate the prediction of
an enhancement of velocity contrasts for shear thinning channelized flows in fractures.
The experimental increase of the front width with the
mean velocity $v$ right above the threshold value
$v_c$  is however slower than the predictions.
The origin of this discrepancy might be investigated by a more refined theory
taking into account the full rheological characteristics of the fluid and aperture variations
along the flow.\\
The theoretical model also allows to reproduce well the experimental front geometry 
for length scales larger than $10\ mm$ which represents the correlation length of the
aperture field in the direction perpendicular to the flow channels. Future work
should investigate the influence of transverse velocity gradients on the shape of the
front for different types of fluids.\\
The results obtained in the present work demonstrate therefore clearly that approaches
developed to analyze channelized Newtonian flows in fractures can be generalized 
to non Newtonian fluids and allow to predict, for instance, the variation of the velocity
contrasts with the rheology.\\
Numerical studies  in $2D$ networks~\cite{Shah1995,Fadili2002} had 
similarly shown  that the flow of shear thinning fluids is localized in a 
smaller number of preferential paths than for Newtonian ones. It has been
suggested that these effects might account for the permeability enhancement
for such fluids mentioned in the introduction: the results obtained in the present
paper may therefore be usefully applicable to the numerical simulation of non Newtonian
flows in fracture networks.\\
A number of questions remain however open and need to be considered in future studies.
 First, the present experiments have been realized with a mean flow parallel to the 
 channels created by the relative shift of the wall surfaces. It will be important to compare
 these results with the case of flow perpendicular to these channels: velocity fluctuations
 in the directions parallel and perpendicular to the flow should then be significantly 
 different from those in the present experiments. Eq.~\ref{eq1} should, for instance, 
 be modified.
Also, the present experiments  deal
 with relatively short path lengths such that transverse 
exchange between channels may be considered as negligible: the results obtained
may therefore be different for longer path lengths. It is also possible that the spatial 
correlation of the velocity field will eventually decay at very long distances although 
this has not been observed in our experimental model.
\section{Acknowledgements}
We are indebted to G. Chauvin and R. Pidoux for their assistance in
the realization of the experimental setup. This work was funded by
E.E.C. through the STREP Pilot plant program SES6-CI-2003-502706 and
by the CNRS-PNRH program. This research  was also supported by a
CNRS-CONICET Collaborative Research Grant (PICS CNRS $2178$), by the
ECOS A03-E02 program and by the I029 UBACyT programs.

\end{document}